# From Understanding to Resonance: A Case Study of Quantum Music and Embodied Science Communication

Kenji Noda (1), Akira Koga (1), Stefan Heusler (2), Koji Hashimoto (3)

(1) Center for the Strategy of Emergence, The Japan Research Institute, Limited, Tokyo 141-0022, Japan

(2) Institut für Didaktik der Physik, Wilhelm-Klemm-Str. 10, 48149 Münster, Germany

(3) Department of Physics, Kyoto University, Kyoto 606-8502, Japan

## Abstract

One challenge in science communication is how to engage audiences with highly abstract scientific fields often perceived as distant, technical, or inaccessible. This Practice Insight examines two initiatives implemented during the International Year of Quantum Science and Technology: Quantum Fest in Japan and Quantum100 in Germany, which together attracted more than 4000 participants. Both employed immersive artistic experiences, including music, visual art, and embodied participation to create alternative entry points into quantum physics. Drawing on post-event surveys, participant comments, and stakeholder interviews, this paper identifies three dimensions of resonance: sensory immersion and reduced psychological barriers, collective embodied meaning-making, and emerging transformative engagement. The findings suggest that resonance-driven approaches can complement explanation-centred science communication by enabling audiences to form meaningful relationships with science before, alongside, or beyond conceptual understanding.

## 1. Introduction

One central goal of science communication is to raise interest in STEM topics (Science, Technology, Engineering, and Mathematics). This goal is particularly challenging in highly abstract domains such as quantum physics, where non-specialist publics may perceive the subject as distant, mathematical, and inaccessible. A key practical question for science communication is therefore how meaningful entry points to such abstract science can be created before, alongside, or beyond full conceptual understanding.

Research on STEM education suggests that interest is shaped not only by subject matter, but also by how science is encountered. The person-object theory of interest (POI; Krapp, 2002a, 2002b) emphasizes the relationship between individuals and objects of interest, while recent work on empathizing approaches to physics education shows that human-centred contexts, emotional perspectives, and social interaction can enhance students' interest in physics [Baron-Cohen, 2009; Baron-Cohen and Wheelwright, 2004; Baron-Cohen et al., 2003; Welberg et al., 2025]. These findings resonate with broader discussions in science outreach, where emotions are increasingly understood as central to how people engage with, understand, remember, trust, or act upon science [Massarani et al., 2025].

This perspective is especially relevant for quantum physics. Affective responses such as wonder and awe, as well as immersive experiences, may provide alternative entry points for publics who do not initially identify with physics or mathematics [Keltner and Haidt, 2003]. Importantly, people may grasp what quantum physics does to society, and why it matters, without first understanding quantum physics in a formal or mathematical sense.

This was the starting point for two entangled international science outreach projects in Japan and Germany, implemented as major events within the International year of Quantum Science and Technology. Together, these initiatives attracted more than 4,000 participants and employed immersive, music-based experiences to communicate quantum physics. Both projects centred on the concert titled 'Fundamental Interactions', which translated concepts from quantum physics into music and visual art while adapting the event design to the specific venues, institutions, and cultural contexts of Japan and Germany.

This paper presents a qualitative case study of these two concrete science communication initiatives. Its purpose is not to measure the statistical effects of the events or to propose a universally replicable model, but to develop practice-oriented insights into the conditions under which affective and embodied engagement with science became possible. Drawing on participant surveys, stakeholder interviews, and related data, the analysis identifies recurring patterns, practical tensions, and limitations within the cases.

Based on these insights, this study develops a resonance-driven framework for science communication practice organized around three interrelated dimensions: sensory immersion and reduced psychological barriers, collective embodied meaning-making, and emerging transformative engagement. Rather than replacing cognitive understanding, the framework positions affective, embodied, and relational engagement as complementary entry points into abstract science. Its contribution lies not in offering a formula for replication, but in articulating transferable design principles for inclusive science communication practices.

## 2. Case Description: Quantum Fest in Japan and Quantum 100 in Germany

### 2.1 Hybrid Event Design in Japan and Germany

This project was implemented in Japan and Germany as part of the UNESCO International Year of Quantum Science and Technology. Despite contextual differences, both events shared a common core: an immersive concert titled *Fundamental Interactions*, which translated concepts from quantum physics into music and visual art through collaboration between composer and director Yannick Paget, theoretical physicist Koji Hashimoto, and the artist collective N'SO KYOTO.

The Japanese case provides the primary empirical basis for identifying recurring patterns across the three dimensions of resonance, whereas the German case offers complementary insights into collective participation, historical framing, and the socio-cultural dimensions of science communication.

**Quantum Fest. Japan**

Quantum Fest was a free public event held on 14–15 June 2025 at Miraikan, Tokyo, as part of the International Year of Quantum Science and Technology. The event was organized by the Physical Society of Japan (JPS), with Takahiro Yamamoto serving as Chair of the Organizing Committee and Koji Hashimoto as Chair of the Programme Committee. The Japan Society of Applied Physics, N'SO KYOTO, Miraikan, and Academist participated as principal partners [Physical Society of Japan, 2025a; Physical Society of Japan, 2025b]. Approximately 1,000 people participated across the two days.

The programme combined public lectures on quantum computing, cosmology, cryptography, and spintronics; guided tours of museum exhibitions; a dialogue on quantum physics and art; and the immersive audiovisual concert *Fundamental Interactions*. The event was also notable as the JPS's first crowdfunding initiative. Its Academist campaign raised ¥9,616,055 from 354 supporters—147% of its ¥6.5 million target—and supported the event and its digital documentation [Academist, 2025]. Public communication formed an integral part of the project: the JPS organized four promotional YouTube Live programmes before the festival and a post-event broadcast featuring physicists, science communicators, and educational creators [Physical Society of Japan, 2025b]. The crowdfunding page also published endorsements from numerous scientists and science communicators, including the JPS president, presenting the festival as an attempt to connect physics, music, and public culture [Academist, 2025]. The post-event survey analysed in this study comprised 98 responses.

**Quantum 100. Germany**

Quantum100 was held at Halle Münsterland on 15 November 2025 as Germany's nationwide closing event for the International Year of Quantum Science and Technology, under the patronage of the German Physical Society. The event was locally organized by Stefan Heusler, Professor at the Institute for Physics Education at the University of Münster, who also served as a scientific advisor to the concert, together with Koji Hashimoto [University of Münster, 2025a; Quantum100, 2025]. Approximately 3,000 people visited the event, while its organizational team comprised more than 400 people; around 250 participated in the concert production alone [University of Münster, 2025b]. The day-long programme included workshops for school groups, twelve public talks, and an exhibition containing approximately thirty contributions from universities, research organizations, and industry. Its culmination was an expanded performance of *Fundamental Interactions*, involving approximately sixty orchestral musicians and 160 choristers, immersive sound, live-generated video, and ceramic percussion [Quantum100, 2025]. The fifth movement introduced the Quantum100 Chorus, whose members wore fluorescent shirts displaying portraits of one hundred physicists. Lyrics by Chris Mosdell transformed Oppenheimer's destructive imagery into a message of life, discovery, and hope [Quantum100, 2025]. The event followed the signing on 14 November of the joint German–Japanese "Declaration for the Future" by the presidents of the German Physical Society (DPG) and JPS at the City Hall of Münster, a historical site associated with the Peace of Westphalia of 1648, linking quantum science with scientific responsibility and nuclear disarmament [University of Münster, 2025c]. The post-event survey analysed in this study comprised 241 responses.

### 2.2 'Fundamental Interactions' as Immersive Artistic Experience

The concert consisted of five movements — I. Electromagnetism, II. Strong Force, III. Weak Force, IV. Gravity, and V. Unity | Quantum100 — in which physical phenomena were translated into sound and visual art. A 360-degree speaker system created an immersive acoustic environment, while real-time generative visuals supported multi-sensory engagement with scientific concepts beyond purely cognitive comprehension.

### 2.3 Integration of Embodiment and Engagement: The Quantum Chorus

A distinctive feature of the German programme was the Quantum Chorus, performed by more than 150 local high-school students aged 14–18 (Figure 1). Each participant wore a T-shirt featuring one of 100 physicists selected for their contributions to quantum physics (Figure 2). For each scientist, participants were provided with a short biography and an accessible explanation of their key contribution, written at a high-school level (in German).

By inviting students to embody these scientific figures during the performance, the project transformed the history of quantum physics from an abstract narrative into a personal and participatory experience. In doing so, it fostered an embodied connection to the scientific community and its historical development.

In the concluding segment, the choir presented a reinterpretation of a well-known quotation associated with J. Robert Oppenheimer, originally derived from the Bhagavad Gita: "Now I am become death, the destroyer of worlds." The quotation gained renewed public visibility through C. Nolan's film *Oppenheimer*. This was rephrased by C. Mosdell, who wrote the text for the Quantum chorus as "Now we've become the breath of life, the discoverers of worlds," thereby transforming a historical narrative of destruction into one of collective responsibility and hope.

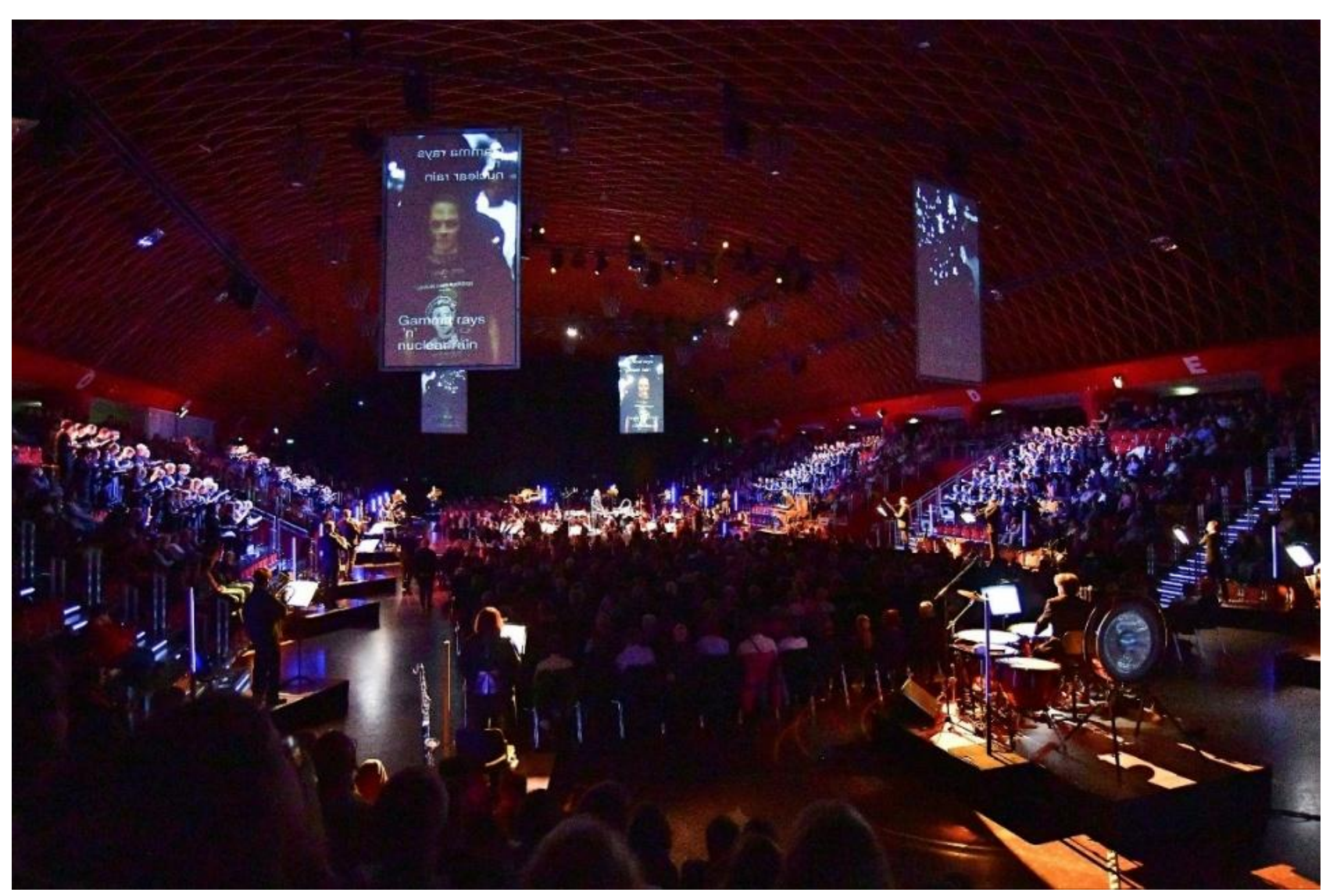

Figure 1. Concert scene at Halle Münsterland, with the Quantum Chorus arranged in four distinct locations throughout the venue, creating an immersive spatial sound experience [Quantum100, 2025].

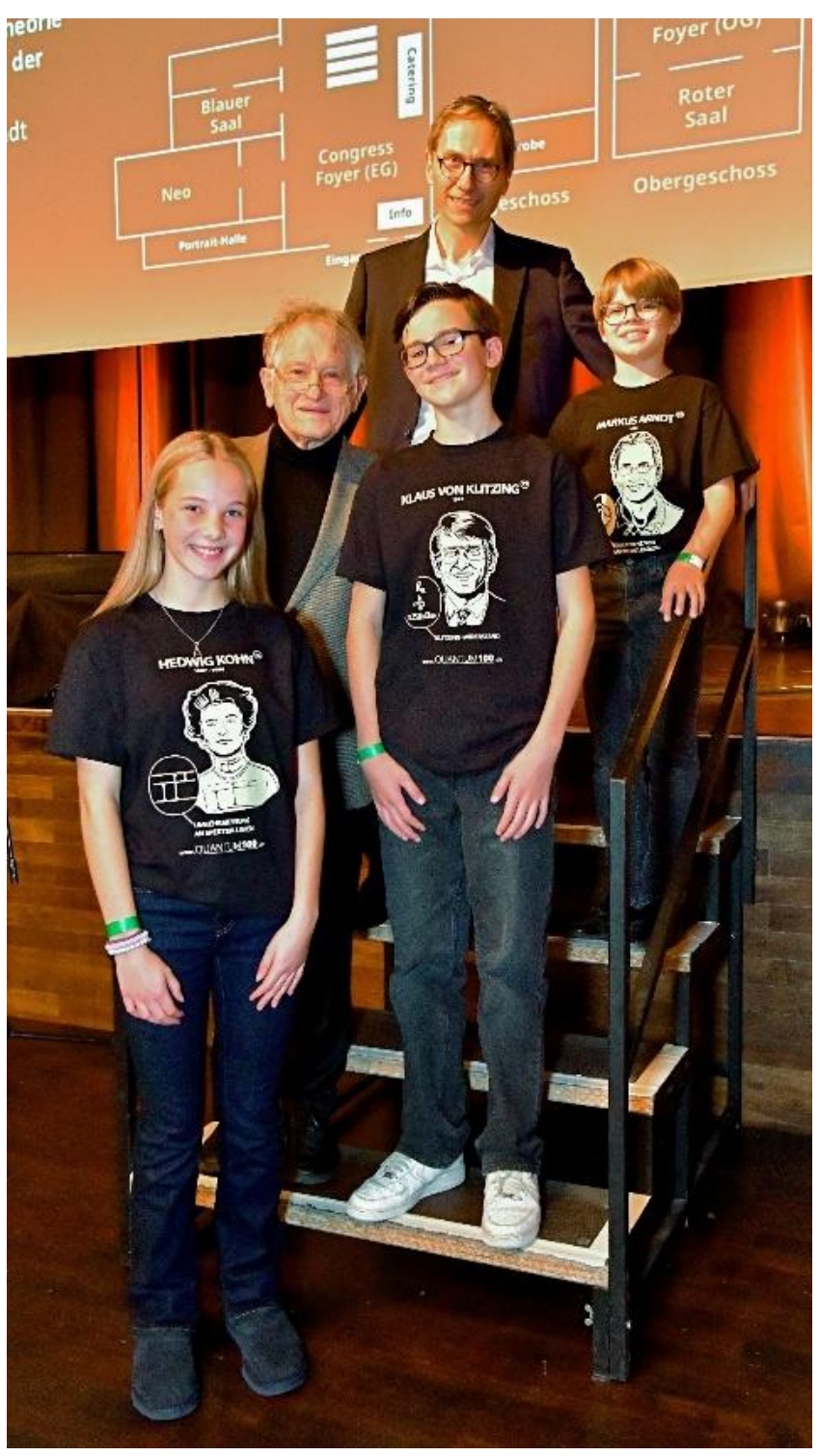

Figure 2. Members of the Quantum Chorus wearing portrait T-shirts representing prominent physicists. Some scientists portrayed in the chorus were present at the event and met their student "representatives."

Although the programme structures differed between Japan and Germany, both cases shared a consistent strategy: facilitating engagement with science not through prior cognitive understanding, but through sensory, affective, and embodied experiences that precede or accompany comprehension. The significance of these cases lies not in their full reproducibility as event formats, but in the way they reveal how science communication can emerge from situated configurations of scientific expertise, artistic practice, affective experience, embodied participation, and local cultural context.

## 3. Methods

This study adopts a qualitative, practice-based approach, drawing on post-event participant surveys, open-ended comments, and semi-structured interviews with key stakeholders involved in the design and implementation of the events. The aim is not to produce a standardized or statistically generalizable intervention model, but to identify recurring patterns and articulate evidence-informed lessons for science communication practitioners and researchers.

Representative comments were selected to illustrate recurring patterns, analytically relevant examples, and tensions or limitations observed in the data. The analysis focused on three dimensions introduced above:

**(1) Sensory immersion and reduced psychological barriers**, including embodied experiences of sound, reduced resistance to scientific language and formulas, and sensory engagement with abstract physics.

**(2) Collective embodied meaning-making**, including scientist–artist co-creation, participant autonomy, relational encounters with scientists, and the collaborative construction of meaning within the event environment.

**(3) Emerging transformative engagement**, including increased curiosity, reconsideration of how science can be expressed, and perceived potential for applying music-based science communication across scientific fields.

## 4. Results

### 4.1 Quantum Fest in Japan

This section presents qualitative findings from the post-event survey conducted after the Quantum Fest in Japan (N = 98), supplemented by interviews with key stakeholders involved in the event. Representative participant responses are summarized in Table 1. The alphanumeric codes following quotations (e.g., I89, L79, P56) refer to anonymized participant IDs assigned to individual survey responses. Additional representative participant responses are provided in Supplementary Table S1.

#### 4.1.1 Sensory Immersion and Reduced Psychological Barriers towards Quantum Physics

At the level of individual experience, many participants described the 360-degree sound environment as a bodily experience that went beyond ordinary music listening. One participant reported physically feeling the sound throughout the venue (L55), while another associated the musical experience with wave-like and particle-like qualities (P56).

The event also appeared to reduce psychological barriers associated with scientific formulas and technical language. Participants who had previously felt distant from physics or mathematical formulas reported that equations, which would normally evoke resistance, were received within the artistic context as expressive forms carrying meaning. One participant described this shift as being able to perceive formulas for the first time as “words carrying a message” (I89). This response indicates that artistic mediation allowed mathematical formulas to be received not merely as technical symbols, but as meaningful forms expressing something about the structure of the world.

Similarly, another participant wrote, “Although it is difficult to approach the essence of physics, I felt that I could glimpse that essence through music” (L79). This comment suggests that sensory immersion provided an affective and aesthetic form of access before full conceptual comprehension. The significance of such responses lies not in demonstrating understanding, but in showing that participants could feel a form of proximity to physics through sound, bodily sensation, and artistic expression.

This interpretation was also supported by editors from a major Japanese popular science magazine, who contrasted the event with more explanation-oriented forms of science communication and described it as a memorable “festival” experience that appeared to attract some audiences beyond the usual science-interested public.

At the same time, some participants still struggled to understand the relationship between quantum physics and music. One wrote, “The performance itself was wonderful, but I didn’t quite get how quantum mechanics and music were connected” (M7), while another stated, “I couldn't understand it myself” (L22). These responses show that sensory immersion may lower psychological barriers, but does not automatically produce conceptual understanding.

#### 4.1.2 Collective Embodied Meaning-Making

Beyond individual sensory experience, the Quantum Fest functioned as a space of collective embodied meaning-making. Participants’ relationships with quantum physics were shaped through interactions among scientists, artists, music, venue design, and their own modes of engagement.

Stakeholder interviews highlighted the collaboration between theoretical physicist Koji Hashimoto and composer-director Yannick Paget. Rather than using music merely to illustrate scientific content, the collaboration allowed quantum mechanics to function as a creative motif, producing a co-created practice in which scientific and artistic forms shaped each other.

The festival format also supported participant autonomy by allowing audiences to move between lectures, exhibitions, and performances at their own pace. One stakeholder involved in science exhibition production observed that the event was not structured as a conventional educational presentation in which audiences passively receive information. Instead, participants could engage with different elements of the festival according to their own interests and curiosity.

This relational dimension was also visible in participants' responses to scientists themselves. Participants did not respond only to scientific themes, but also to the visible passion, enthusiasm, and affection that the physicists expressed toward their research. Responses such as "I became interested in people who approach physics with such love and passion" (N28) and "I felt as though I could glimpse inside the minds of physicists" (I72) suggest that the event reduced the perceived distance between scientists and participants. Participants also generated their own interpretations of scientific and social meaning. For example, one participant interpreted the second movement, "Strong Interaction," as "a requiem for lives lost in the nuclear age" (P59). This response shows that the physical concept was not received only as scientific content, but was reconfigured through sound, personal emotion, and historical memory.

#### 4.1.3 Emerging Transformative Engagement: Curiosity and Future-Oriented Reflections

The survey and interview data do not allow claims about long-term impact or stable transformation. However, some responses provide hints that the event stimulated curiosity, prompted some participants to reconsider how physics can be expressed, and raised expectations for future science-art practices.

Some participants reported increased curiosity toward quantum physics and related scientific themes. One participant in Tokyo wrote, "I wanted to know more about this world, which is made up of elementary particles" (I27). This suggests that affective and sensory engagement may stimulate further curiosity even in the absence of full conceptual understanding.

Other responses indicate that the event prompted some participants to reconsider how physics itself can be expressed. One participant noted that the event let them feel that physical phenomena could be expressed not only through mathematics but also through music (L93). The participant further noted that this challenged their previous assumption that mathematics was the only language for describing physical reality. Rather than reflecting increased scientific knowledge, this response suggests a shift in how physics was imagined and interpreted.

Some participants also reflected on the broader potential of the event format. One participant expressed interest in experiencing similar music-based approaches in other scientific fields (O41), while another suggested that such collaborations could foster new forms of innovation across both physics and music (O93). These comments indicate that participants saw possibilities extending beyond the specific case of quantum physics.

Taken together, these responses suggest that the event generated more than immediate emotional reactions. While the data do not demonstrate long-term transformative outcomes, they indicate that affective and embodied engagement may encourage curiosity, reflection, and imagination about new relationships between science, art, and society. Immersive science-art formats may therefore make quantum physics experientially approachable, while also revealing the limits of affective access: resonance may precede understanding, but it does not guarantee conceptual comprehension.

| Category | Case | Context | ID | Representative Response |
| --- | --- | --- | --- | --- |
| Sensory immersion and reduced psychological barriers | Japan | Reasons and comments: “Quantum and Art” lecture | I89 | As someone who had long felt distant from physics and mathematical formulas, I was able, for the first time, to perceive formulas as words carrying a message. |
| | Japan | Potential of music–quantum physics integration | M7 | The performance itself was wonderful, but I didn’t quite get how quantum mechanics and music were connected. |
| | Germany | Audience feedback on Quantum Chorus | Q121 | What a wonderful idea to include a choir of high-schoolers. Their fluorescing T-shirts made their appearance all the more radiant. |
| Collective embodied meaning-making | Japan | Reasons for deepened motivation to learn quantum physics | N28 | I became interested in people who approach physics with such love and passion. |
| | Germany | Audience feedback on Quantum Chorus | Q52 | Overall, the collaborative spirit embodied in the concert conveyed a powerful and important message. |
| | Germany | Audience feedback on Quantum Chorus | Q56 | The enthusiasm of the choir was truly exceptional. They showed how young people can be inspired and engaged. |
| Emerging transformative engagement | Japan | Reasons and comments: “Quantum and Art” lecture | I27 | I felt a desire to learn more about this world made of elementary particles. I want to be able to talk more with my child about their favorite subject, physics. |
| | Germany | Audience feedback on Quantum Chorus | Q15 | The event was accessible to newcomers, diverse yet focused, and emphasized contemporary societal issues. |

Table 1. Illustrative Evidence from the Japanese and German Cases Across the Three Dimensions of Resonance

Notes

- IDs correspond to anonymized participant responses collected in the post-event survey.
- Responses were translated from Japanese and German into English by the authors.
- Categories were developed through thematic analysis and correspond to the three analytical dimensions of the resonance-driven framework: sensory immersion and reduced psychological barriers, collective embodied meaning-making, and emerging transformative engagement.
- Minor edits were made for readability while preserving the original meaning.

### 4.2 “Quantum100” in Germany

This section presents qualitative findings from the post-event survey conducted after Quantum100 in Germany (N = 241) with spectators of the event, and TV interviews in nationwide television (ZDF, “Volle Kanne”) with members of the Quantum Chorus. Overall, audience feedback was overwhelmingly positive: more than 80% of responses were positive, approximately 10% were neutral, and around 10% contained critical remarks. Criticism mainly concerned individual lectures, organizational aspects, or personal preferences regarding the musical style. Given the experimental nature of the composition and the challenge of engaging with quantum physics through contemporary music, a fully unanimous reception would neither be expected nor necessarily desirable.

One audience, for example, described the concept as “wonderful” and expressed interest in seeing it repeated, while also noting that some movements seemed more closely aligned with the event’s concept than others and that an inclusion policy was not clearly communicated. This response illustrates a nuanced audience reception that combined appreciation with constructive criticism.

Overall, the findings were similar to those from the Japanese case (see Table. 1). This section therefore focuses on the impact of the Quantum Chorus, a unique feature of the event in Germany that especially stimulated embodied meaning-making and emerging transformative engagement (see Section 2.3 and Table. 1). One 17-year-old chorus member stated in a TV-interview:

*“When I hear the term quantum physics, I tend to think of it as something very abstract and far removed from everyday life. But when you start to consider all the things it is used for, you realize that it is actually very much a part of our daily lives”*

Indeed, the relevance of quantum physics can be appreciated even without prior content knowledge. In many science outreach initiatives, affective approaches are used to stimulate interest, yet cognitive understanding is still implicitly assumed to be the primary gateway to engagement. Participation is therefore often designed around the comprehension, interpretation, and discussion of scientific concepts. In contrast, our approach emphasizes physics as a socio-scientific endeavor and highlights its impact on society. Presenting portraits of 100 physicists on the choir members' T-shirts encouraged participants to identify not only with the science but also with the people behind it. A booklet containing short biographies of all 100 physicists, written in German, was distributed to support this engagement.

## 5. Discussion

### 5.1 From understanding to resonance

The two cases suggest that communication of highly abstract science need not begin with formal explanation. Participants encountered quantum physics through sound, space, bodily sensation, artistic mediation, and social interaction. These encounters did not replace cognitive understanding, but created conditions under which quantum physics could be experienced as meaningful and approachable without requiring full conceptual understanding.

This shift is described here as a movement from understanding-centred communication toward resonance-driven communication. In an understanding-centred model, success is primarily evaluated by the accurate transmission of scientific content. A resonance-driven approach instead asks how people form relationships with science through affect, embodiment, identification, and shared experience, and how such relationships may later support curiosity and learning.

Here, resonance refers to the formation of a meaningful relationship with science through which science becomes experienced as personally relevant, relatable and connected to one’s own life.

### 5.2 Three dimensions of resonance

The three dimensions of resonance clarify what can be transferred from these situated cases. First, sensory immersion and reduced psychological barriers show that abstract science can become approachable through bodily and aesthetic experience. Sound, vibration, visual mediation, and the aesthetic reframing of equations allowed some participants to feel proximity to quantum physics without reporting conceptual mastery. However, comments expressing confusion about the link between quantum mechanics and music show that sensory access needs interpretive support.

Second, collective embodied meaning-making highlights the relational nature of science communication. In Japan, this was evident in scientist–artist co-creation and in participants’ responses to the visible passion of scientists. In Germany, the Quantum Chorus intensified this dimension by turning high-school students into embodied representatives of physicists, making scientific history performative and relational.

Third, emerging transformative engagement should be understood cautiously. The data do not demonstrate long-term transformation, but they show signs of changed orientation:

curiosity, reconsideration of how physics can be expressed, interest in science-art formats for other fields, and reflections on the ethical and historical significance of physics. Transformation therefore appears as an emerging possibility rather than as a measured outcome.

### 5.3 Implications for inclusive science communication

The practical value of these cases does not lie in reproducing the exact event format. Both depended on situated conditions, including scientist–artist collaboration, immersive concert design, symbolic venues, Münster's historical association with peace, and the Quantum Chorus. What can be transferred are design principles: creating affective and bodily access before full understanding; treating artistic collaboration as genuine co-creation rather than illustration; giving participants autonomy; making scientists visible as passionate and relatable people; and connecting scientific topics to local cultural, ethical, or historical meanings.

For inclusive science communication, these cases suggest that explanation remains important but should not be the only gateway into science. Audiences who feel distant from physics, mathematics, or expert culture may nevertheless form meaningful relationships with science through music, narrative, embodiment, identification, and shared experience.

### 5.4 Limitations and future research

This study has several limitations. First, the findings are based primarily on post-event surveys, open-ended comments, and stakeholder interviews. These data are suitable for identifying recurring patterns and developing practice-based insights, but they do not allow causal claims about the effects of the events. Second, the study does not measure long-term changes in participants' knowledge, attitudes, or behavior. The term "transformative engagement" should therefore be understood as indicating emerging possibilities rather than confirmed long-term transformation.

Third, the Japanese and German cases differed in scale, structure, funding model, modes of participation, and available data. The Japanese event was partly made possible through crowdfunding, whereas the German event was funded mainly by foundations, with ticket sales accounting for only a small share of the budget. These different funding and participation structures may have shaped how audiences entered into relation with the events before attending, including their initial expectations, sense of involvement, or motivation to participate. However, the available data do not allow us to assess these effects systematically. The Japanese analysis draws heavily on participant survey comments and stakeholder interviews, while the German case emphasizes the distinctive role of the Quantum Chorus and related audience and media responses.

Future research could examine how affective and embodied engagement develops over time, whether such experiences lead to sustained interest in science, and how similar approaches might be adapted to other scientific fields. Comparative studies across different cultural contexts and artistic media would also help refine the transferable design principles proposed here.

## 6. Conclusion

This paper examined Quantum Fest in Japan and Quantum100 in Germany as two qualitative Practice Insight cases of embodied and music-based science communication. The analysis shows how abstract science can become publicly meaningful not only through explanation, but also through sensory immersion, artistic mediation, social encounter, and participatory embodiment.

The proposed resonance-driven vocabulary does not offer a universal model or a replicable recipe. Rather, it identifies practice-relevant conditions under which diverse publics may form a first relationship with science before, alongside, or beyond conceptual understanding. For quantum physics and other abstract domains, such conditions may be crucial for making science communication more inclusive, culturally situated, and experientially meaningful, with a focus on science as a socio-scientific issue.

**Acknowledgments**

We would like to express our sincere gratitude to all the individuals and organizations whose commitment made Quantum Fest in Japan and Quantum100 in Germany possible. For Quantum Fest, we are particularly grateful to Takahiro Yamamoto, Chair of the Organizing Committee, for his leadership, and to the Physical Society of Japan, the Japan Society of Applied Physics, Miraikan – The National Museum of Emerging Science and Innovation, N'SO KYOTO, and Academist Inc. for their organizational, institutional, and operational support. We also warmly thank the 354 crowdfunding supporters, whose contributions enabled the festival and its digital documentation, as well as the scientists, science communicators, sponsors, volunteers, and YouTube guests who helped communicate and promote the project. For Quantum100, we thank the German Physical Society, the University of Münster, the local organizing team, EIN Quantum NRW, and Halle Münsterland. We are grateful for the support of the Wilhelm and Else Heraeus Foundation, the Sparkasse Münsterland Ost Foundation, Universitätsgesellschaft Münster, the Klaus Tschira Foundation. We also thank the Münster Student Orchestra, the choirs of Gymnasium Paulinum, the choir Piano 22/30, and all participating artists, scientists, exhibitors, students, technical personnel, and volunteers. Finally, we thank the audiences and survey respondents in Japan and Germany, whose participation and reflections made this study possible.

The authors used ChatGPT (GPT-5.5 Thinking, OpenAI) to support language editing, drafting, and stylistic revision during manuscript preparation. The authors reviewed, revised, and take full responsibility for the final content.